\documentclass[fleqn,twoside]{article}
\usepackage{espcrc2}
\usepackage{graphicx}

\newcommand{\AmS}{{\protect\the\textfont2
  A\kern-.1667em\lower.5ex\hbox{M}\kern-.125emS}}
\title{Finite band effects over the linear conductance through a quantum wire coupled with a quantum dot:
X-boson treatment}
\author{R. Franco\address{Departamento de
F\'{\i}sica, Pontificia Universidade Cat\'olica do Rio de Janeiro (PUC-Rio),
22452-970, Caixa Postal: 38071  Rio de Janeiro, Brazil}\thanks{e-mail: rfranco@fis.puc-rio.br}
, E. V. Anda\addressmark
\hspace{0.1cm} and M. S. Figueira\address{Instituto de F\'{i}sica Universidade Federal Fluminense (UFF). Avenida litor\^{a}nea s/n, 
24210-340, Caixa Postal: 100.093, Niter\'{o}i, Rio de Janeiro, Brazil}} %\thanks{Work supported by National Research Council (CNPq),
\begin{document}

\date{\today}

\begin{abstract}
We study the electronic transport through a quantum wire (QW) with a strong side coupled quantum dot (QD). 
We obtain a linear conductance with lateral peaks when the 
gate voltage, is located near the edge of the conduction band. The calculated density of states shows, 
that these peaks are associated with renormalized localized levels out side of the conduction band. These results are compatibles
with recent experimental results. \hspace{0.5cm}$Key$ $words:$  Quantum Dot; Fano resonance; Mesoscopic system; X-boson.
\vspace{1pc} 
\end{abstract}

\maketitle

In a previous work, we apply the X-boson method \cite{X-boson2} for the single impurity case to
describe the transport problem through a quantum wire (QW) with a side coupled quantum dot (QD),
in the limit when the coulomb interaction at the QD was $U\rightarrow \infty$ \cite{QDnosso}. 
In this work we apply the X-boson method to the same system studied in our previous work \cite{QDnosso}, 
but considering a stronger coupling between the QD and  the QWW, in a similar situation described in a recent 
experimental work \cite{Gores2000}.

The model we use to describe the system is the Anderson impurity Hamiltonian
in the $U \rightarrow \infty $ limit using the Hubbard operators representation, given by %\cite{Hubbard4}. 

\begin{eqnarray}
H &=&\sum_{\mathbf{k},\sigma }E_{\mathbf{k},\sigma }c_{\mathbf{k},\sigma
}^{\dagger }c_{\mathbf{k},\sigma }+\sum_{\sigma }\ E_{f,\sigma }X_{f,\sigma
\sigma }\\ \nonumber
&&+ \sum_{\mathbf{k},\sigma}\left( V_{f,\mathbf{k},\sigma }X_{f,0\sigma
}^{\dagger }c_{\mathbf{k},\sigma }+V_{f,\mathbf{k},\sigma }^{\ast }c_{%
\mathbf{k},\sigma }^{\dagger }X_{f,0\sigma }\right) .  \label{Eq.3}
\end{eqnarray}

\noindent The first term represents the
conduction electrons ($c$-electrons), associated with the wire. The second
describes the QD and the last one corresponds to the interaction
between the $c$-electrons and the QD. 
At low temperature and small bias voltage, the electronic  transport is coherent and a
linear-conductance is obtained using the Landauer-type formula \cite{QDnosso}

\begin{equation}
G=\frac{2e^2}{h}\int{\left(-\frac{\partial n_{F} }{\partial \omega}%
\right)S(\omega) d\omega} ,  \label{Landauer}
\end{equation}

\noindent where $n_{F}$ is the Fermi distribution function and $S(\omega )$
is the transmission probability of an electron with energy $\hbar \omega $, given
by $S(\omega )=\Gamma ^{2}|{G_{00}^{\sigma}}|^{2}$.  $V$ is the matrix element connecting the QD with its nearest
site, belonging to the wire, represented by the label $0$, $\Gamma =V^{2}/\Delta $, with $\Delta =%
\frac{\pi V^{2}}{2W}$,  $W$ is the half-width conduction band and $G_{00}^{\sigma}(\omega )$ is the dressed Green
function (GF) at the wire site $0$. This function can be written in terms of the GF at the QD, $G_{qd}^{\sigma}$ and the GF that
describes the conduction electrons $g_{c}^{\sigma}(z)=-\frac{1}{2W}ln\left| \frac{z+W}{z-W}\right| $ (ballistic channel), as 
$G_{00}^{\sigma}=(g_{c}^{\sigma }V)^{2}G_{qd}^{\sigma }+g_{c}^{\sigma }$.

\begin{figure}[ht]
\begin{center}
\includegraphics[clip,width=0.45\textwidth, height=0.25\textheight, angle=0.0]{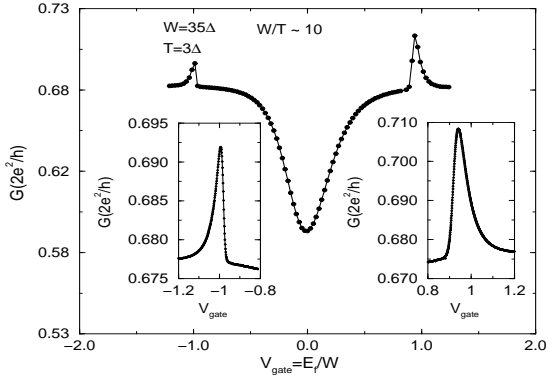}
\caption{Conductance $G$ vs $V_{gate}=E_{f}/W$. The maximum value of the linear conductance was taken from the maximum constant value 
(the ``background'') for the linear 
conductance in the experimental result of G\"{o}res et. al. \cite{Gores2000} in their Fig. 4-a. The insets show in detail the lateral 
resonances} %, a form of take in account the 
%contribution of no magnetic impurities, that we don't considered in our computation.}
\label{fig1_QD}
\end{center}
\end{figure}

Using the chain X-boson method, and considering a constant density of states for the wire, $-W\leq \varepsilon _{%
\mathbf{k}}\leq W,$ the GF for the QD is given by \cite{X-boson2,QDnosso}

\begin{equation}
G_{qd}^{\sigma }(z)=\frac{-D_{\sigma }}{z-\tilde{E_{f}}-\frac{V^{2}D_{\sigma
}}{2W}ln\left| \frac{z+W}{z-W}\right| },  \label{Gff}
\end{equation}

\noindent where $z=\omega +i\eta $, the quantity $D_{\sigma }=\left\langle
X_{0,0}\right\rangle +n_{f,\sigma }$ is responsible for the correlation in
the chain X-boson approach \cite{X-boson2} and 
$\tilde{E_{f}}=E_{f}+\Lambda $, where $\Lambda $ is a parameter of the X-boson 
method for the impurity case \cite{X-boson2}.
In the Fig. \ref{fig1_QD} we show the linear conductance $G$ vs $V_{gate}$, associated with the energy $E_{f}$ of the localized
state (QD) in the experimental works \cite{Gores2000}, for a half-width band $W=35\Delta$, with 
$\Delta=\frac{\pi V^{2}}{2W}$. %We considered in a previous work \cite{QDnosso} the case $W=100\Delta$, that is, we are increasing 
%the coupling between the QWW and the QD. 
It is possible to see the lateral resonances, when $V_{gate}=E_{f}$ is located near the edge 
of the conduction band, associated with the QW. Our numerical calculations show that these lateral peaks are absent 
for low temperatures.  
\label{Sec3}

\begin{figure}[ht]
\begin{center}
\includegraphics[clip,width=0.3352\textwidth, height=0.33\textheight, angle=-90.0]{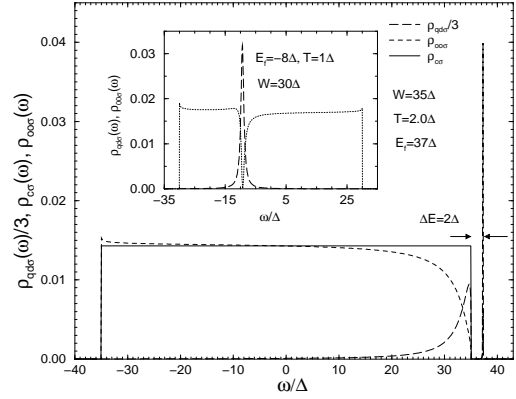}
\caption{Density of states associated with the Green's functions of the system.}
\label{fig2_QD}
\end{center}
\end{figure}

In the Fig. \ref{fig2_QD} we present the density of states associated with the wire site $0$, $\rho_{oo,\sigma}(\omega)$, and at the QD 
$\rho_{qd,\sigma}(\omega)$ respectively; the density of states $\rho_{c,\sigma}(\omega)$, associated with the undressed GF 
$g_{c}^{\sigma}(\omega )$, for a value of $V_{gate}=E_{f}>+W$ is also shown. It is possible to see the appearance of a localized level out of the 
conduction band. A similar level was obtained at the left side of the border of the conduction band. The inset present the same results, but for 
a value of $V_{gate}=E_{f}$ inside of the conduction band.

We obtain similar results to the measurents presented in the experimental work of G\"{o}res et. al \cite{Gores2000} for the case of a 
single electron transistor (Fig 2a in \cite{Gores2000}). In this work a small magnetic field ($\vec{H}$) was applied at a constant 
temperature, increasing one of the lateral peaks (Fig 6 in \cite{Gores2000}). We expect that the presence of $\vec{H}$ 
reduces the separation between the localized level and the nearest edge of the band for electrons with spin oriented in the opposited direction
of $\vec{H}$, increasing the transport in 
the channel associated with this spin orientation. This process could exponentially increase the height of the peak associated with this 
conduction channel when $\vec{H}$ is increased.

We acknowledge the financial support of the National Research Council (CNPq),
Latin American Center of Physics (CLAF) and Rio de Janeiro State Research Foundation (FAPERJ).

\end{document}